# 3D-Printed Optics for Wafer-Scale Probing

Mareike Trappen [1, 2, a], Matthias Blaicher[1, 2], Philipp-Immanuel Dietrich[1, 2, 3], Tobias Hoose[1, 2], Yilin Xu[1, 2], Muhammad Rodlin Billah[1, 2], Wolfgang Freude[1] and Christian Koos [1,2,3,b]

[1] Institute of Photonics and Quantum Electronics (IPQ), Karlsruhe Institute of Technology (KIT), Engesserstraße 5, 76131 Karlsruhe, Germany
[2] Institute for Microstructure Technology (IMT), Karlsruhe Institute of Technology (KIT), Hermann-von-Helmholtz-Platz 1, 76344 Eggenstein-Leopoldshafen, Germany
[3] Vanguard Photonics GmbH, Hermann-von-Helmholtz-Platz 1, 76344 Eggenstein-Leopoldshafen, Germany
[a] mareike.trappen@kit.edu, [b] christian.koos@kit.edu

**Abstract** *Mass production of photonic integrated circuits requires high-throughput wafer-level testing. We demonstrate that optical probes equipped with 3D-printed elements allow for efficient coupling of light to etched facets of nanophotonic waveguides. The technique is widely applicable to different integration platforms.*

**Introduction**

Scalable production of photonic integrated circuits (PIC) requires high-throughput wafer-level testing to evaluate device performance early in the fabrication process, prior to entering costly and complex chip separation and packaging steps. To this end, light must be efficiently coupled from optical probes to in-plane waveguides with high reproducibility. Presently, optical wafer-scale probing predominantly relies on grating structures[1,2] to couple light from in-plane nanophotonic waveguides to single-mode fibers (SMF) that approach the chip from an essentially surface-normal direction. While such grating couplers (GC) have been widely used for silicon photonic[2] (SiP), or silicon nitride (SiN) waveguides[3], the concept is inherently limited to typical 3-dB bandwidths of 50 nm and cannot be transferred to other photonic integration platforms such as indium phosphide (InP) or low index-contrast silica waveguides. In addition, future PIC designs increasingly rely on broadband edge-coupling[4,5] (EC) via vertical waveguide facets, which can be fabricated on wafer level by deep etching of dicing trenches into the wafer surface. Coupling to and from vertical waveguide facets represents a major challenge of wafer-level testing, which was recently addressed by optical probes that exploit total internal reflection at a polished angled facet of a planar lightwave circuit (PLC)[6]. In this experiment coupling to planar SiN waveguides with comparatively large mode field diameter (MFD) of approximately 10 µm was demonstrated, leading to a total insertion loss of 5.7 dB per coupling interface. Transferring this approach to probing of high index-contrast waveguides with smaller cross sections is challenging, since the PLC probe always emits a divergent beam, which does not match the mode field of small in-plane waveguides.

In this paper we present a novel approach to optical probes, that allows for low-loss coupling between etched facets of in-plane waveguides and standard single-mode fibers approaching the wafer from a surface-normal direction. The concept relies on *in-situ* 3D-printing of periscope-like freeform coupling structures to the SMF endfaces. These SMF-lens-periscopes (SMF-LP) combine a total internal reflection (TIR) mirror for redirecting the light with a refractive optical surface for focusing the emitted beam. SMF-LP stand out due to their small size, which easily fits into standard dicing trenches and allow to adapt the mode fields to a wide variety of in-plane waveguide cross sections. We demonstrate the viability of the concept by probing of waveguides on the SiN (TriPleX) and SiP platform, which feature mode field sizes of 10 µm and 1.5 µm and lead to coupling losses of 3.1 dB and 5.7 dB, respectively.

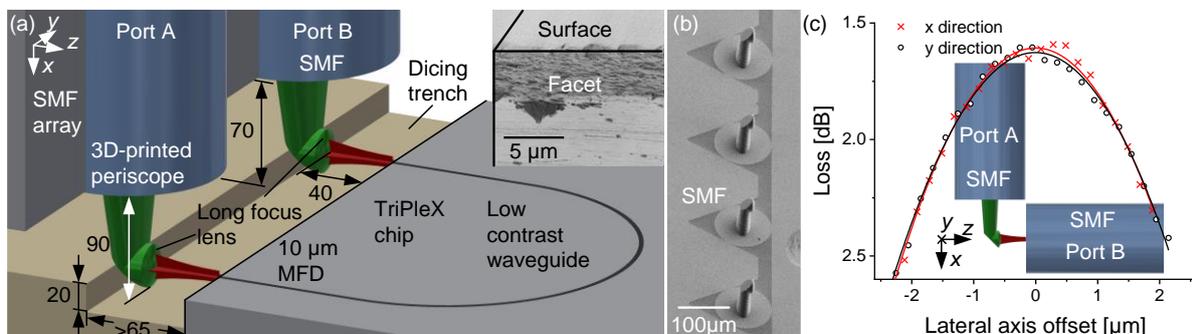

**Fig. 1:** Wafer-level probing of a SiN waveguide using a standard single-mode fiber (SMF) array with 3D-printed "lensed periscopes" (SMF-LP, working distance 50 µm), which fit into a standard dicing trench. (a) Schematic of probing arrangement for a loop-back transmission measurement with a typical mode field size of 10 µm. All dimensions are indicated in µm. **Inset:** SEM image of TriPleX chip facet. (b) Scanning electron microscope (SEM) image of 3D-printed SMF-LP. (c) Coupling efficiency between a SMF-LP and test SMF as a function of lateral offset.

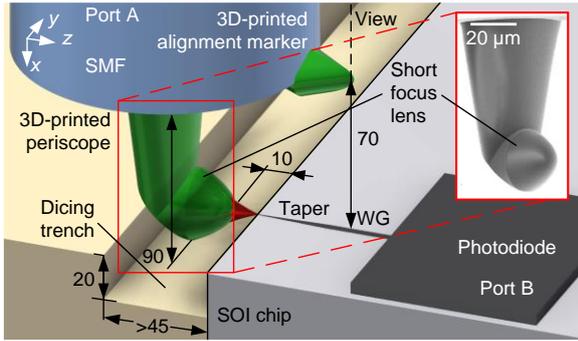

**Fig. 2:** Wafer-level probing of a silicon photonic (SiP) waveguide using a standard SMF array with 3D-printed SMF-LP (working distance 10 µm). The alignment markers help in positioning the SMF array relative to the on-chip waveguides. The chip waveguide is terminated with a Ge photodiode. For clarity, only one SMF is shown although the SMF-LP were fabricated on a V-groove array of 8 fibers. **Inset:** SEM image of 3D-printed periscope. All dimensions are indicated in µm.

**Probing Concept**
The concept of wafer-level probing is illustrated in Fig. 1 (a). The optical probes access the vertical facets of on-chip waveguides from top via the chip edges or the dicing trenches. For efficient coupling, a vertically positioned array of SMF is supplemented by a TIR mirror to redirect the beam horizontally, Fig. 1(a). A focusing lens allows to adapt the spot size of the emitted beam to the mode field of the on-chip waveguide. We exploit direct-write two-photon laser lithography for *in-situ* structuring of the SMF-LP at the SMF end face with high alignment accuracy in respect to the fiber core[7], Fig. 1(b). The height of the fiber array endface over the wafer surface is chosen to be 70 µm to prevent mechanical damage while aligning the probe. The position and the shape of the lens surface is chosen to match the mode field of the waveguide facet while maintaining a working distance of at least 10 µm between the apex of the lens and the vertical facet. The SMF-LP fit into standard dicing trenches with minimum accessible trench widths of 65 µm for the TriPleX interface, Fig. 1, and 45 µm for the SiP interface, Fig. 2. The minimum trench depth (20 µm) is determined by the lens diameter. To simplify alignment of the probe with respect to the on-chip waveguides, we print alignment markers to end-faces of the SMF array, Fig. 2, which can be used for automated alignment by machine vision in future implementations. The marker tips point along the SMF-LP optical axis and allow to coarsely align the probe, before active alignment techniques are used for refined positioning.

**Fabrication**
The SMF-LP structures were *in-situ* printed into the commercially available photoresist IP-Dip using a customised 3D two-photon lithography system (Nanoscribe Photonic Professional GT, 40× objective lens with NA = 1.4), which is complemented by a proprietary high precision automated alignment software. Slicing and hatching writing distances were both set to 100 nm. The mode field diameter amounts to approximately 1.5 µm (10 µm) for the SiP (TriPleX) waveguide. For optimum coupling to these mode fields, the aspherical lens surfaces are optimised using either the physical-optics module of Zemax OpticStudio (TriPleX) or an in-house wave-propagation[8] software (SiP).

**Demonstrators**
To prove the viability of the concept, we designed two different SMF-LP probes for coupling to TriPleX and SiP waveguides at a wavelength of 1550 nm. Both SMF-LP were positioned at the edge of a single photonic chip or in front of a SMF by use of a 6 axis alignment stage.
For characterizing the TriPleX SMF-LP, we first measure the coupling to a test SMF as a function of lateral offset, see Fig. 1(c) and Tab. 1, Experiment 1. For an SMF optimally aligned with respect to the SMF-LP we find a coupling loss of 1.6 dB between the input of the SMF-LP-equipped fiber ("port A") and the output of the test SMF ("port B"). The 1-dB lateral alignment tolerance amounts to ± 2.3 µm. The excess loss of 1.6 dB is attributed to the loss of the TIR reflection mirror, to Fresnel reflections at the lens surface and to fabrication imperfections, with significant potential for further improvement[7]. For estimation of the coupling efficiency to the TriPleX chip, we use a loop-back waveguide with 250 µm spacing between the input and output facet and measure the fiber-to-fiber loss between to be 6.2 dB, see Fig. 1(a) and Tab.1, Experiment 2. Assuming identical coupling interfaces and a lossless waveguide loop, the attenuation per coupling amounts to 3.1 dB. For comparison we measured the fiber-to-fiber loss for coupling into the TriPleX chip with a SMF array, see Tab. 1, Experiment 3, leading to an overall loss of 10.3 dB. Assuming again the waveguide to be lossless we calculate an attenuation of 5.1 dB per coupling interface. The high coupling loss are attributed to the high surface roughness of the chips facet, see Inset of Fig. 1 (a). Note that, in contrast to previous demonstrations[6], the coupling losses of the SMF-LP-based optical probe are smaller than those obtained for direct coupling to SMF. We attribute this to the ability of the SMF-LP to focus the light to a distant working spot (working distance: 40 µm), hence allowing to obtain best coupling efficiency without bringing the facet into direct physical contact, which always bears the risk of damaging the optical surface. In contrast to this, the standard SMF array in Experiment 3 needs to be in direct contact with the etched surface.

**Tab. 1:** Overview of coupling experiments and measured losses. "Loss A to B" and "Loss A to C" denote the transmission losses between the respective reference points A, B, and C as indicated in the second column.

| Experiment no. | Schematic of Coupling Experiment | Loss A to B | Loss A to C |
|---|---|---|---|
| 1 | A◯—[SMF][LP]—[SMF]—◯B | 1.6 dB | - |
| 2 | A◯—[SMF][LP]—[TriPleX]—[LP][SMF]—◯B | 6.2 dB | 3.1 dB |
| 3 | A◯—[SMF]—[TriPleX]—[SMF]—◯B | 10.3 dB | 5.1 dB |
| 4 | A◯—[SMF][LP]—[SiP][PD]—◯B | 5.7 dB | - |
| 5 | A◯—[GC]—[SiP]—[GC]—◯B | 9.0 dB | 4.5 dB |

For characterising the SiP SMF-LP we measured the coupling to a tapered on-chip waveguide featuring a MFD of about 1.5 µm, see Fig. 2 and Tab.1, Experiment 4. The waveguides were fabricated through the SiP foundry process offered by the A*Start Institute of Microelectronics (IME) in Singapore. The waveguide was terminated by an on-chip Ge photodiode, which was electrically contacted to measure the photocurrent. Neglecting the on-chip propagation losses, the attenuation loss attributed to the SMF-LP interface is 5.7 dB. In the experiment, the photodiode responsivity was determined using another photodiode, nominally identical to the one used in the coupling experiment, which was connected to waveguides terminated by GC. Moreover, we determine the GC losses by measuring the transmission through a SiP reference structure, Tab. 1, Experiment 5. The used SIP chips show GC losses of 4.5 dB. Note that direct interfacing of a standard SMF to a SiP device facet would not lead to conclusive results due to the large MFD mismatch. This emphasizes once more the importance of shaping the mode field by a refractive surface of the SMF-LP.

In comparison, the TriPleX SMF-LP show lower loss than the SIP SMF-LP. This result was expected from simulations, for which the TriPleX SMF-LP showed an optimized coupling efficiency of 99% (10 µm MFD) compared to 69% (1.6 dB) for the SIP SMF-LP (1.5 µm MFD). Note that these simulations do not account for Fresnel reflection at the lens surface, which causes additional loss. Still, these findings leave room for further improving the coupling efficiency of the optical probe.

## Summary

We have introduced and experimentally demonstrated a technique for high-precision wafer-scale probing of edge-coupled PIC. The concept relies on compact 3D-printed coupling structures, which can be introduced into standard dicing trenches etched into the wafer surface. We demonstrate losses of 3.1 dB and 5.7 dB when coupling to silicon nitride (TriPleX SiN) and silicon photonic on-chip waveguides. The concept is applicable to SMF arrays, which allow for simultaneous readout of multiple channels and therefore enable parallel measurements. The fabrication process of direct laser writing allows high flexibility in designing the coupling elements, making the concept applicable to a wide variety of optical integration platforms.

## Acknowledgements

We thank K. Wörhoff and A. Leinse, both at LioniX BV, for providing TriPleX chips. This work was supported by the Bundesministerium für Bildung und Forschung (BMBF) joint project PRIMA (13N14630), the Helmholtz International Research School for Teratronics (HIRST), the European Research Council (ERC Starting Grant 'EnTeraPIC', # 280145; ERC Consolidator Grant 'TeraSHAPE', # 773248), by the H2020 Photonic Packaging Pilot Line PIXAPP (# 731954), by the Deutsche Forschungsgemeinschaft (DFG) through CRC WavePhenomena (# 1173), and by the Karlsruhe Nano-Micro Facility (KNMF).